\documentclass[a4paper,12pt]{article}
\pdfoutput=1 

\usepackage{jcappub} 
\usepackage{bm}
\usepackage{soul}
\usepackage{indentfirst}
\usepackage{amsmath}
\usepackage{graphicx}
\usepackage{float}
\usepackage{amssymb}
\usepackage{subfigure}
\usepackage{amssymb}
\usepackage{psfrag}
\usepackage{hyperref}
\usepackage{epstopdf}
\usepackage{tensor}
\usepackage{comment}
\usepackage[symbol]{footmisc}
\hypersetup{
  colorlinks=true,
  linkcolor=blue,
  citecolor=blue,
}

\allowdisplaybreaks[1]

\begin{document}

\title{Model-independent reconstruction of the primordial curvature power spectrum from  PTA  data}

\author{Zhu~Yi,$^{a}$}
\author{Zhi-Qiang~You,\note{Corresponding author.}$^{b,*}$}
\author{and You~Wu$^{c,d,*}$}

\affiliation{$^a$Advanced Institute of Natural Sciences, Beijing Normal University, Zhuhai 519087, China}
\affiliation{$^b$Henan Academy of Sciences, Zhengzhou 450046, Henan, China}
\affiliation{$^c$College of Mathematics and Physics, Hunan University of Arts and Science, Changde, 415000, China}
\affiliation{$^d$Department of Physics and Synergistic Innovation Center for Quantum Effects and Applications, Hunan Normal University, Changsha, Hunan 410081, China}

\emailAdd{yz@bnu.edu.cn}
\emailAdd{you\_zhiqiang@whu.edu.cn}
\emailAdd{youwuphy@gmail.com}

\abstract{
Recently released data from pulsar timing array (PTA) collaborations provide strong evidence for a stochastic signal consistent with a gravitational-wave background, potentially originating from scalar-induced gravitational waves (SIGWs). However, in order to determine whether the SIGWs with a specific power spectrum of curvature perturbations can account for the PTA signal, one needs to estimate the energy density of the SIGWs, which can be computationally expensive.
In this paper, we use a model-independent approach to reconstruct the primordial curvature power spectrum using a free spectrum cross over from $10^{1}\,\mathrm{Mpc}^{-1}$ to $10^{20}\,\mathrm{Mpc}^{-1}$ with NANOGrav 15-year data set. 
Our results can simplify the task of assessing whether a given primordial curvature power spectrum can adequately explain the observed PTA signal without calculating the energy density of SIGWs.
}

\maketitle

\section{Introduction}
The detection  of a common-spectrum process  demonstrating the Hellings-Downs angular correlation feature inherent in gravitational waves (GWs) has been reported by the North American Nanohertz Observatory for Gravitational Waves (NANOGrav) \cite{NANOGrav:2023gor, NANOGrav:2023hde}, 
Parkers Pulsar Timing Array (PPTA) \cite{Zic:2023gta, Reardon:2023gzh},  
European Pulsar Timing Array (EPTA) \cite{Antoniadis:2023lym, Antoniadis:2023ott}, 
and Chinese Pulsar Timing Array (CPTA) \cite{Xu:2023wog}, recently.   
Assuming the source to be an ensemble of binary supermassive black hole inspirals and adopting a fiducial $f^{-2/3}$ characteristic-strain spectrum, its estimated strain amplitude is approximately $\sim 10^{-15}$ at a reference frequency of $1 ~ \rm{yr}^{-1}$  \cite{NANOGrav:2023gor,Reardon:2023gzh,Antoniadis:2023ott,Xu:2023wog}.  
However, the precise source of this intriguing signal, whether attributed to supermassive black hole binaries (SMBHBs) or other cosmological phenomena, remains a subject of active investigation ~\cite{NANOGrav:2023hvm,Antoniadis:2023xlr,Yi:2023mbm, Franciolini:2023pbf,Liu:2023ymk,Vagnozzi:2023lwo,Cai:2023dls,Bi:2023tib,Wu:2023hsa,Franciolini:2023wjm,You:2023rmn,Jin:2023wri,Liu:2023pau,An:2023jxf, Zhang:2023nrs, Das:2023nmm,Balaji:2023ehk,Du:2023qvj,Oikonomou:2023qfz,Yi:2023npi,Chen:2023zkb,Liu:2023hpw}. 
Among the various hypotheses, a particularly promising candidate is the scalar-induced gravitational waves (SIGWs); and this paper focuses on investigating the scenario where the observed signal arises from SIGWs.
For other physical processes contributing to the PTA band, please see \cite{Zhu:2018lif,Li:2019vlb,Chen:2021wdo,Wu:2021kmd,Chen:2021ncc,Chen:2022azo,PPTA:2022eul,IPTA:2023ero,Wu:2023pbt,Wu:2023dnp,InternationalPulsarTimingArray:2023mzf,Wu:2023rib,Bi:2023ewq,Chen:2023uiz}. 

SIGWs, accompanied by the formation of primordial black holes (PBHs), originate from primordial curvature perturbations generated during the inflationary epoch  \cite{Hawking:1971ei,Carr:1974nx,Ananda:2006af,Baumann:2007zm,Saito:2008jc,Alabidi:2012ex,Sasaki:2018dmp,Nakama:2016gzw,Kohri:2018awv,Di:2017ndc,Cheng:2018yyr,Cai:2019amo,Cai:2018dig,Cai:2019elf,Cai:2019bmk,Wu:2020drm,Cai:2020fnq,Pi:2020otn,Domenech:2020kqm,Liu:2018ess,Liu:2019rnx,Liu:2020cds,Liu:2021jnw,Liu:2022iuf,Chen:2019xse,Yuan:2019fwv,Yuan:2019wwo,Yuan:2019udt,Carr:2020gox,Liu:2020vsy,Liu:2020bag,Papanikolaou:2021uhe,Papanikolaou:2022hkg,Chakraborty:2022mwu,Chen:2018czv,Chen:2018rzo,Chen:2019irf,Chen:2021nxo,Chen:2022fda,Liu:2022wtq,Zheng:2022wqo,Chen:2022qvg,Meng:2022low,Garcia-Saenz:2023zue}. 
The requirement for significant SIGWs is a power spectrum amplitude for primordial curvature perturbations, approximately $\mathcal{A}_{\zeta}\sim \mathcal{O}(0.01)$, which is seven orders of magnitude larger than constraints from cosmic microwave background (CMB) anisotropy observations, $\mathcal{A}_{\zeta}= 2.1\times 10^{-9}$ \cite{Planck:2018jri}. 
This rapid enhancement in the primordial curvature power spectrum can be accomplished through inflation models with a transitional ultra-slow-roll phase \cite{Martin:2012pe, Motohashi:2014ppa,Yi:2017mxs, Garcia-Bellido:2017mdw, Germani:2017bcs, Motohashi:2017kbs,Ezquiaga:2017fvi, Gong:2017qlj, Ballesteros:2018wlw,Dalianis:2018frf, Bezrukov:2017dyv,Kannike:2017bxn,Lin:2020goi,Lin:2021vwc,Gao:2020tsa,Gao:2021vxb,Yi:2020kmq,Yi:2020cut,Yi:2021lxc,Yi:2022anu,Zhang:2020uek,Pi:2017gih,Kamenshchik:2018sig,Fu:2019ttf,Fu:2019vqc,Dalianis:2019vit,Gundhi:2020zvb,Cheong:2019vzl,Zhang:2021rqs, Zhang:2021vak,Kawai:2021edk,Cai:2021wzd,Chen:2021nio,Zheng:2021vda,Karam:2022nym,Ashoorioon:2019xqc}.   
Detectable not only by PTAs but also by space-based GW detectors such as LISA~\cite{Danzmann:1997hm, Audley:2017drz}, Taiji \cite{Hu:2017mde}, TianQin \cite{Luo:2015ght, Gong:2021gvw}, and DECIGO~\cite{Kawamura:2011zz}, SIGWs provide invaluable insights into the physics of the early Universe. For instance, through the observational data of SIGWs, constraints on primordial curvature perturbations at small scales can be obtained \cite{Inomata:2018epa, Yi:2022ymw}.

In exploring the possibility of  SIGWs as an explanation for the PTA signal, the primordial curvature power spectrum is commonly characterized through special models, such as the $\delta$-function, lognormal form, or broken power law parameterizations \cite{NANOGrav:2023hvm, You:2023rmn}. By employing these parameterizations and using the PTA data, one can derive constraints on the primordial curvature power spectrum at small scales.  However, a drawback of this approach lies in its dependency on the specific parameterizations of the curvature power spectrum.  To overcome this limitation, in this paper, we adopt a linear interpolation parameterization with a large number of bins.  The linear interpolation parameterization is versatile enough to depict various types of primordial curvature power spectra, as long as the number of bins is sufficiently large.  
By employing linear interpolation parameterization, researchers can analyze and interpret data in a more general way, allowing for the exploration of a broader range of possibilities and reducing the influence of preconceived assumptions about the underlying functional form of the power spectrum.  
This near-model-independent methodology can provide more robust constraints on the primordial curvature power spectrum. By this parameterization, we reconstruct the primordial curvature power spectrum, and the amplitude is displayed in a free-spectrum framework in this paper.

The organization of this paper is as follows. 
Section \ref{sec:sigw} gives the energy density of the SIGWs, and section \ref{sec:nbox} introduces the model-independent parameterization of the primordial curvature power spectrum. In Section \ref{sec:result}, we present our results, and the conclusions are drawn in Section \ref{sec:con}.

\section{The scalar-induced gravitational waves}\label{sec:sigw}
If the scalar perturbations seeded from the primordial curvature perturbations generated during inflation are large enough, the secondary order of the linear scalar perturbations can act as a source to induce GWs.  In this section, we give the energy density of SIGWs during radiation domination in detail.   The perturbed metric in the Newtonian gauge can be expressed as 
\begin{equation}
ds^2=a^2(\eta)\left[-(1+2\Phi)d\eta^2+\left\{(1-2\Phi)\delta_{ij}+\frac{1}{2}h_{ij}\right\}dx^i x^j\right],
\end{equation}
where ${\Phi}$ represents the Bardeen potential, and ${h_{ij}}$ is the tensor perturbation following the transverse-traceless gauge condition, ${\partial^{i}h_{ij}=h^{i}_{i}=0}$. In the Fourier space, the tensor perturbation can be represented as 
\begin{equation}
h_{ij}\left(\bm{x},\eta\right)=\int\frac{d^3\bm{k}}{\left(2\pi\right)^{3/2}}\text{e}^{i\bm{k}\cdot\bm{x}}\left[h^{+}_{\bm{k}}\left(\eta\right)e^{+}_{ij}\left(\bm{k}\right)
+h^{\times}_{\bm{k}}\left(\eta\right)e^{\times}_{ij}\left(\bm{k}\right)\right].
\end{equation}
Here, ${e^{+}_{ij}\left(\bm{k}\right)}$ and ${e^{\times}_{ij}\left(\bm{k}\right)}$ represent the plus and cross-polarization tensors,
\begin{equation}
\begin{split}
 &e^{+}_{ij}\left(\bm{k}\right)=\frac{1}{\sqrt{2}}\left[e_i\left(\bm{k}\right)e_j\left(\bm{k}\right)
 -\overline{e}_i\left(\bm{k}\right)\overline{e}_j\left(\bm{k}\right)\right],\\& 
 e^{\times}_{ij}\left(\bm{k}\right)=\frac{1}{\sqrt{2}}\left[e_i\left(\bm{k}\right)
 \overline{e}_j\left(\bm{k}\right)+\overline{e}_i\left(\bm{k}\right)e_j\left(\bm{k}\right)\right],
 \end{split}
\end{equation}
and ${e_{i}\left(\bm{k}\right)}$ and ${\overline{e}_{i}\left(\bm{k}\right)}$ are orthonormal basis vectors orthogonal to the wave vector $\bm{k}$, $\bm{e}\cdot\bm{\overline{e}} =\bm{e}\cdot\bm{k}=\bm{\overline{e}}\cdot\bm{k}=0$.

Disregarding anisotropic stress, the equation of motion for tensor perturbations in Fourier space, for either polarization, takes the form
\begin{equation}\label{E4}
h^{''}_{\bm{k}}\left(\eta\right)+2\mathcal{H}h^{'}_{\bm{k}}\left(\eta\right)+k^2h^{}_{\bm{k}}\left(\eta\right)=4S_{\bm{k}}\left(\eta\right),
\end{equation}
where $\eta$ is the conformal time, $d\eta = dt/a$, a prime denotes the derivative with respect to the conformal time,   ${\mathcal{H}=a'/a}$ represents  the conformal Hubble parameter, and the source term is given by
\begin{equation}
\label{hksource}
 S_{\bm{k}}=\int \frac{d^3\tilde{k}}{(2\pi)^{3/2}}e_{ij}(\bm{k})\tilde{k}^i\tilde{k}^j
 \left[2\Phi_{\tilde{\bm{k}}}\Phi_{\bm{k}-\tilde{\bm{k}}} \phantom{\frac{1}{2}}+ 
 \frac{1}{\mathcal{H}^2} \left(\Phi'_{\tilde{\bm{k}}}+\mathcal{H}\Phi_{\tilde{\bm{k}}}\right)
 \left(\Phi'_{\bm{k}-\tilde{\bm{k}}}+\mathcal{H}\Phi_{\bm{k}-\tilde{\bm{k}}}\right)\right],
 \end{equation}
where  $\Phi_{\bm{k}}$  is the  Bardeen potential in the Fourier space. By using  the  primordial curvature  perturbations $\zeta_{\bm{k}}$ generated in the inflation, the Bardeen potential can be expressed as
 \begin{equation}\label{phik}
 \Phi_{\bm{k}}=\frac{3+3w}{5+3w}T(k,\eta) \zeta_{\bm{k}},
 \end{equation}
 where  $T(k,\eta)$ is  the transfer function,
  \begin{equation}\label{transfer}
 T(k,\eta)=3\left[\frac{\sin\left(k\eta/\sqrt{3}\right)-\left(k\eta/\sqrt{3}\right)
 		\cos\left(k\eta/\sqrt{3}\right)}{\left({k\eta}/{\sqrt{3}}\right)^3}\right].
 \end{equation}
The solution for $h_{\bm{k}}\left(\eta\right)$ in equation \eqref{E4} can be obtained through the Green's function method,  
\begin{equation}\label{E7}
h_{\bm{k}}\left(\eta\right)=\frac{4}{a\left(\eta\right)}\int^{\eta}d\overline{\eta}G_{k}\left(\eta,\overline{\eta}\right)
a\left(\overline{\eta}\right)
S_{\bm{k}}\left(\overline{\eta}\right),
\end{equation}
where the Green function is 
 \begin{equation}
    G_{\bm{k}}\left(\eta,\overline{\eta}\right)=\frac{\sin\left[k\left(\eta-\overline{\eta}\right)\right]}{k}.
\end{equation}
The definition of the power spectrum of SIGWs is  
\begin{equation}\label{Ep}
\langle h_{\bm{k}}\left(\eta\right)h_{\bm{p}}\left(\eta\right)\rangle
=\delta^3\left(\bm{k}+\bm{p}\right) \frac{2\pi^2}{k^3}\mathcal{P}_h\left(k,\eta\right).
\end{equation}
Combining Eqs. \eqref{hksource}-\eqref{Ep}, we get
\cite{Kohri:2018awv,Lu:2019sti,Baumann:2007zm,Inomata:2016rbd,Espinosa:2018eve,Ananda:2006af}
 \begin{equation}\label{Eph2}
 \begin{split}
 \mathcal{P}_h(k,\eta)=
 4\int_{0}^{\infty}\!\!dv\int_{|1-v|}^{1+v}\!du \left[\frac{4v^2-(1-u^2+v^2)^2}{4uv}\right]^2 I_{RD}^2(u,v,x)\mathcal{P}_{\zeta}(kv)\mathcal{P}_{\zeta}(ku),
 \end{split}
 \end{equation}
where $\mathcal{P}_\zeta$ is the primordial curvature power spectrum with the definition,
\begin{equation}
    \langle \zeta_{\bm{k}}\zeta_{\bm{p}} \rangle = \delta^3(\bm{k}+\bm{p}) \frac{2\pi^2}{k^3}\mathcal{P}_\zeta.
\end{equation}
Here, $u=|\bm{k}-\tilde{\bm{k}}|/k$,  $v=\tilde{|\bm{k}|}/k$, $x= k \eta$, and 
 \begin{equation}
 	\label{irdeq1}
 	\begin{split}
 		I_{\text{RD}}(u, v, x)=&\int_0^x dy\, y \sin(x-y)\{3T(uy)T(vy)\\
 		&+y[T(vy)u T'(uy)+v T'(vy) T(uy)]\\
 		&+y^2 u v T'(uy) T'(vy)\},
 	\end{split}
 \end{equation}
where the subscript ``RD" denotes the radiation-dominated era with the equation of state satisfying $w=1/3$.
 The energy density of Gravitational waves is 
 \begin{equation}
 	\label{density}
 	\Omega_{\mathrm{GW}}(k,\eta)=\frac{1}{24}\left(\frac{k}{aH}\right)^2\overline{\mathcal{P}_h(k,\eta)}.
 \end{equation}
Combining  equation \eqref{Eph2} and definition \eqref{density},  we obtain \cite{Espinosa:2018eve,Lu:2019sti}
 \begin{equation}
 	\label{SIGWs:gwres1}
 	\begin{split}
 		\Omega_{\mathrm{GW}}(k,\eta)=&\frac{1}{6}\left(\frac{k}{aH}\right)^2\int_{0}^{\infty}dv\int_{|1-v|}^{1+v}du  \left[\frac{4v^2-(1-u^2+v^2)^2}{4uv}\right]^2 \overline{I_{\text{RD}}^{2}(u, v, x)} \mathcal{P}_{\zeta}(kv)\mathcal{P}_{\zeta}(ku),
 	\end{split}
 \end{equation}
where $\overline{I_{\text{RD}}^{2}}$ is the oscillation time average of $I_{\text{RD}}^{2}$. 
At late time, $k\eta \gg 1$, i.e., $x\rightarrow \infty$, the time average of ${I^{2}_\mathrm{RD}\left(u,v,x\rightarrow\infty \right)}$ is \cite{Kohri:2018awv}
\begin{equation}\label{IRD}
\begin{split}
 \overline{I_{\mathrm{RD}}^2(v,u,x\rightarrow \infty)} &=  \frac{1}{2x^2} \left( \frac{3(u^2+v^2-3)}{4 u^3 v^3 } \right)^2 \bigg\{\pi^2 (u^2+v^2-3)^2 \Theta \left( v+u-\sqrt{3}\right)   \\
&   +  \left( -4uv+(u^2+v^2-3) \ln \left| \frac{3-(u+v)^2}{3-(u-v)^2} \right| \right)^2 \bigg\}.
\end{split}
\end{equation}
The energy density evolution of gravitational waves follows the same way as that of radiation. Using this property, we can easily determine the current energy density of gravitational waves 
 \begin{equation}\label{d}
 	\Omega_{\mathrm{GW}}(k,\eta_0)=\frac{\Omega_{r,0}  \Omega_{\mathrm{GW}}(k,\eta)}{\Omega_{r}(\eta)},
 \end{equation}
where $\Omega_{r,0}$ represents the current energy density of radiation, and $\Omega_{r}(\eta)=1$ during the generation of SIGWs.

\section{The model-independent parameterization}\label{sec:nbox}
In investigating the issue of SIGWs, the primordial curvature power spectra are often characterized using specific forms, where the most frequently employed parameterizations are the $\delta$-function, lognormal form, and broken power law parameterizations. Consequently, the outcomes of such investigations are dependent on these specific parameterizations.  Hence,  the exploration of a model-independent or, at the very least, near-model-independent approach is indeed valuable. Following this principle, we parameterize the primordial curvature power spectrum using a linear interpolation form with a total of $n$  bins.  The linear interpolation parameterization is robust to depict different kinds of primordial curvature power spectra, as long as the number of bins is large enough. While the more bins imply a larger amount of computation, we need to strike a balance between these two contradictions. In the following, we give the precise details of the linear interpolation parameterization of the primordial curvature power spectrum. 

On the larger scales where $k \leq k_\mathrm{large} = 1 ~{\rm Mpc^{-1}}$, the primordial curvature power spectrum is already well-constrained by CMB observational data.  The primordial curvature power spectrum can be effectively characterized using a power-law form,
\begin{equation}
     \mathcal{P}_\zeta (k) = A_{\rm CMB} \left(\frac{k}{k_*}\right)^{n_s-1},
\end{equation}
where, from the CMB observational data, the amplitude is $A_{\rm CMB} = 2.1\times 10^{-9}$, the scalar tilt is 
 $n_s=0.965$, and the pivot scale is  $k_* =0.05 {\rm Mpc^{-1}}$ \cite{Planck:2018jri}. 
 
For the scales where  $k_\mathrm{large} < k \leq k_{\rm end} =10^{20}~{\rm Mpc^{-1}}$, 
we divide the primordial curvature power spectrum into three distinct regions: $k_\mathrm{large}< k<k_{\mathrm{pta1}}$,  $k_{\mathrm{pta1}} \leq k \leq k_{\mathrm{pta2}}$, and $k_{\mathrm{pta2}}<k\leq k_{\rm end}$. We set the scale points $k_\mathrm{pta}$ as follows 
\begin{equation}
   \log_{10}(k_{\mathrm{pta1}}/\mathrm{Mpc}^{-1}) =5, \quad \log_{10}(k_{\mathrm{pta2}}/\mathrm{Mpc}^{-1}) =10,
\end{equation}
ensuring that the region between these scale points adequately covers the  PTA  scales.   The smallest scale $k_\mathrm{end}$ is determined by the $e$-folds of the inflation, $e$-folds $\approx \ln(k_\mathrm{end}/ k_*)$ $ = 50$.  
To implement linear interpolation, selecting appropriate data points is essential. 
For the scales within $k_\mathrm{large}< k <  k_\mathrm{pta1}$, we set the scale points $k_i$ as $\log_{10}(k_{i+1}/\mathrm{Mpc}^{-1}) = \log_{10}(k_{i}/\mathrm{Mpc}^{-1}) + 1$, beginning with  $\log_{10}(k_1/\mathrm{Mpc}^{-1}) = 2$.   For the scales within $k_\mathrm{pta1} \leq k \leq  k_\mathrm{pta2}$,  we set the scale points $k_i$ as $\log_{10}(k_{i+1}/\mathrm{Mpc}^{-1}) = \log_{10}(k_{i}/\mathrm{Mpc}^{-1}) + 1/4$ with the initial point $\log_{10}(k_1/\mathrm{Mpc}^{-1}) = \log_{10}(k_\mathrm{pta1}/\mathrm{Mpc}^{-1})$.   Due to their relevance to PTA scales, additional scale points are chosen for this range. 
Finally, for scales within $k_\mathrm{pta2} < k \leq  k_\mathrm{end}$, we set the scale points $k_i$  as $\log_{10}(k_{i+1}/\mathrm{Mpc}^{-1}) = \log_{10}(k_{i}/\mathrm{Mpc}^{-1}) + 1$ with the initial point $\log_{10}(k_1/\mathrm{Mpc}^{-1}) = \log_{10}(k_\mathrm{pta2}/\mathrm{Mpc}^{-1}) +1$.  These scale points $k_i$ are displayed in Figure \ref{fig:free_Pk}.

Using these chosen scale points $k_i$, for scales where  $k_\mathrm{large} < k \leq k_{\rm end} =10^{20}~{\rm Mpc^{-1}}$, the primordial curvature power spectrum can be represented through the linear interpolation form,
\begin{equation}
\log_{10} \mathcal{P}_\zeta (k) =\log_{10} A_i + \frac{\log_{10} A_{i+1}- \log_{10}A_{i}}{\log_{10} k_{i+1}-\log_{10} k_i} \times(\log_{10} k -\log_{10} k_i), \quad k_i <  k \leq k_{i+1}.
\end{equation}
In conclusion, the  primordial curvature power spectrum is expressed as 
\begin{equation}\label{para:pl}
\log_{10}\mathcal{P}_\zeta (k) = 
\left\{
\begin{aligned}
 &\log_{10} A_{\rm CMB} + (n_s-1)\log_{10}\left(k/k_*\right), \quad   k\leq k_\mathrm{large},\\
&\log_{10} A_i + \frac{\log_{10} A_{i+1}- \log_{10}A_{i}}{\log_{10} k_{i+1}-\log_{10} k_i} \times(\log_{10} k -\log_{10} k_i), \quad k_i  < k\leq k_{i+1}.
\end{aligned}
\right.
\end{equation}


\section{Methodology and Results}\label{sec:result}
\begin{figure}[htbp!] 
\centering
\includegraphics[width=\textwidth]{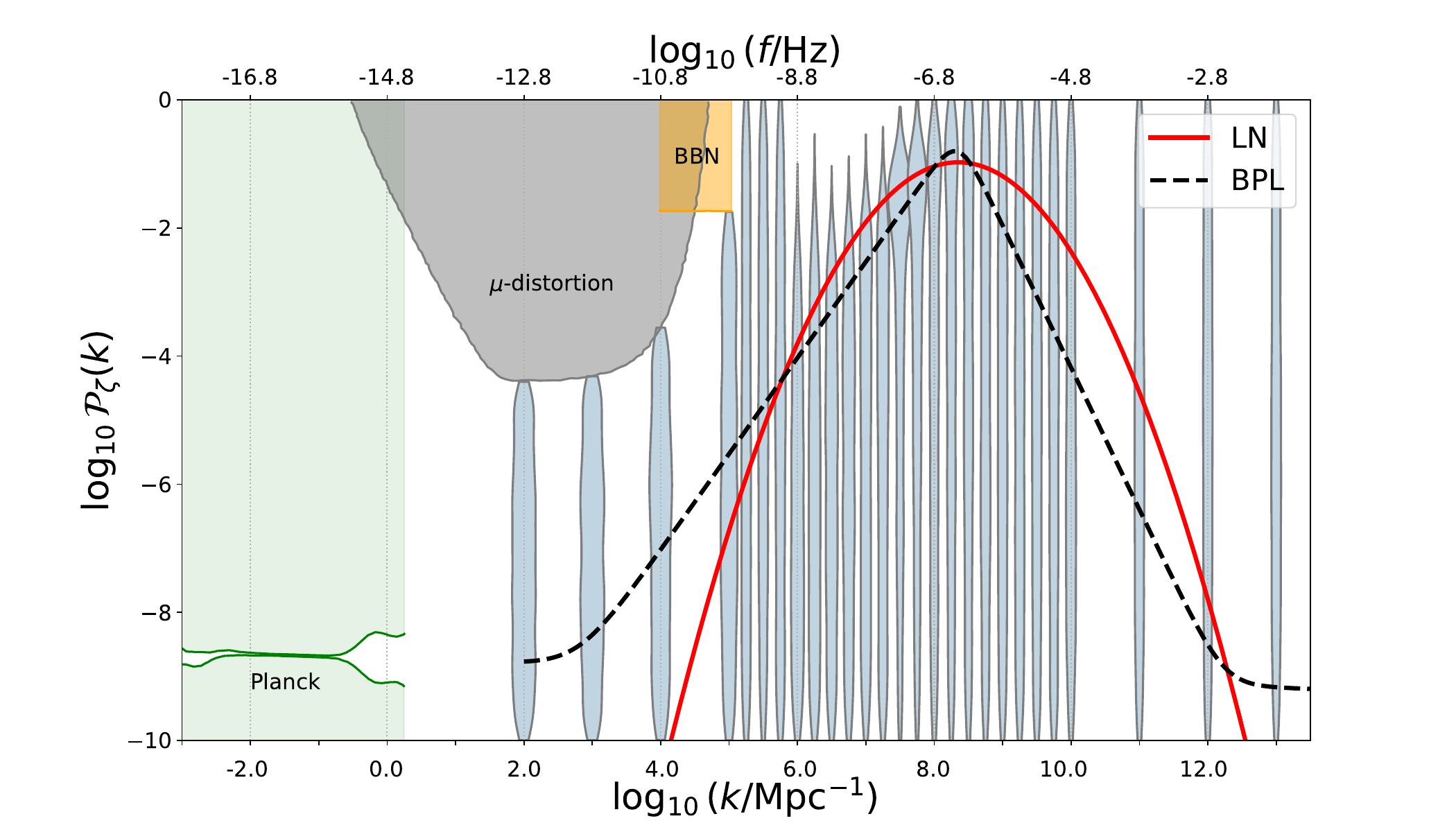}
\caption{The constraints on the primordial curvature power spectrum. The violins are the allowed regions to account for the NANOGrav 15-year data set in a free-spectrum way. The light green shaded region is excluded by the CMB observations \cite{Planck:2018jri}. The orange and gray regions are the constraints from the effect on the ratio between neutron and proton during the Big Bang nucleosynthesis (BBN) \cite{Inomata:2016uip} and $\mu$-distortion of CMB \cite{Fixsen:1996nj}, respectively. The red and black dashed lines are the primordial curvature power spectrum with lognormal (LN) and broken-power-law forms (BPL). 
}
\label{fig:free_Pk}
\end{figure}

In this section, we use a Bayesian analysis on the NANOGrav 15-year data set to reconstruct the amplitude of the primordial curvature power spectrum described by Eq. \eqref{para:pl}. We utilize the 14 frequency bins of the NANOGrav 15-year data set \cite{NANOGrav:2023gor, NANOGrav:2023hvm} to fit the posterior distributions of these amplitudes $A_i$. The analysis was carried out using the \texttt{Bilby} code \cite{Ashton:2018jfp}, employing the \texttt{dynesty} algorithm for nested sampling \cite{NestedSampling} with 1000 live points (${\rm nlive} = 1000$).
The log-likelihood function was obtained by evaluating the energy density of SIGWs at the 14 specific frequency bins. Subsequently, we computed the sum of the logarithms of the probability density functions from 14 independent kernel density estimates \cite{Moore:2021ibq}. Consequently, the likelihood function can be expressed as
\begin{equation}
\ln \mathcal{L}(\mathbf{\Theta})=\sum_{i=1}^{14} \ln \mathcal{L}_i\left(\Omega_{\mathrm{GW}}\left(f_i, \mathbf{\Theta} \right)\right),
\end{equation}
where $\boldsymbol{\Theta}$ represents the collection of parameters in the parameterization given by Eq.  \eqref{para:pl}, 
and the relation of frequency and wave number is 
\begin{equation}
    f =\frac{k}{2\pi} \simeq 1.6~ {\rm nHz} ~ \left(\frac{k}{10^{6}~{\rm Mpc} ^{-1}}\right).
\end{equation}
For  scales  within the range of $2 \leq \log_{10} (k_i / \mathrm{Mpc}^{-1}) \leq 5$, the priors for the amplitude $\log_{10} A_i$  employ  a  uniform distribution denoted as $ \mathrm{U}[-10, \log_{10}(A_c)]$, where $A_c$ is the lower limit of the $\mu$-distortion or BBN constraints at the corresponding scale. For other scales, the prior for each $\log_{10} A_i$ is $\mathrm{U}[-10, 0]$, and posterior distributions are illustrated in Figure \ref{fig:free_Pk}.

Using the linear interpolated model \eqref{para:pl}, the constraints on the continuous primordial curvature power spectrum  are actually on the bin amplitude parameters $A_i$.
The violins in  Figure \ref{fig:free_Pk} illustrate the posteriors of the amplitudes of the primordial curvature power spectrum $A_i$.   
Although the full scales information of the primordial curvature power spectrum is contained in SIGWs as demonstrated by the integration of Eq. \eqref{SIGWs:gwres1},  the NANOGrav 15-year data set can only provide constraints on the primordial curvature power spectrum within a specific range of scales, approximately $k / \mathrm{Mpc}^{-1} \sim [10^{6}, 10^{9}]$. Outside of this scale range, there are few constraints on the amplitudes of the primordial curvature power spectrum.
For the scales $k / \mathrm{Mpc}^{-1} > 13$, the posteriors of the amplitudes $A_i$ are similar with  $k / \mathrm{Mpc}^{-1}= 13$ (almost no constraints on amplitudes), so we don't display them in the Figure \ref{fig:free_Pk}.  In Figure \ref{fig:free_Pk}, the light green shaded region is excluded by the Planck 2018 data \cite{Planck:2018jri}, and the narrow allowed region was derived using Planck TT, TE, EE+lowE+lensing+BK15 data and 12 knots for the cubic spline interpolation  \cite{Planck:2018jri}. Because of the high precision measurements of CMB anisotropies by the Planck mission, the constraints on the primordial curvature power spectrum at large scales are quite strong. The orange and gray regions are the constraints from the effect on the ratio between neutron and proton during the Big Bang nucleosynthesis (BBN) \cite{Inomata:2016uip} and $\mu$-distortion of CMB \cite{Fixsen:1996nj}, respectively.

To assess the quality of our model-independent reconstruction, we compare them with model-dependent scenarios, as illustrated in Figure \ref{fig:free_Pk}. 
The red line and black dashed line represent the primordial curvature power spectrum using the lognormal form and the broken power-law form, respectively.  
These two parameterizations utilize the best-fitted values that account for the NANOGrav 15-year data set via SIGWs \cite{You:2023rmn}. The model-independent reconstruction is consistent with the model-dependent results.  Consequently, using this model-independent reconstruction, we can directly assess whether a given primordial curvature power spectrum can explain the NANOGrav 15-yrs signal via SIGWs, without the need for calculating the corresponding SIGWs explicitly.   
{In Figure \ref{fig:free_Pk},  only two bins can be well constrained, while others are weakly constrained,} making it difficult to distinguish different parameterizations of the primordial curvature power spectrum.
This is consistent with the previous results that the NANOGrav 15-year data set can not effectively differentiate between various parameterizations of the primordial curvature power spectrum, as discussed in Ref. \cite{NANOGrav:2023hvm, You:2023rmn}.

\section{Conclusion}\label{sec:con}
The signal detected by NANOGrav, PPTA, EPTA, and CPTA collaborations recently can be explained by SIGWs.  
However, when investigating the plausibility of SIGWs as an explanation for the PTA signals, the primordial curvature power spectrum often needs to be parameterized in specific forms. 
As a result, the conclusions drawn depend on the special models.  
In this paper, we parameterize the primordial curvature power spectrum by using a linear interpolation method. 
This parameterization is near-model-independent for it can depict a wide array of possible primordial curvature power spectra, as long as the number of bins is large enough.  
By adopting this linear interpolation parameterization and treating the NANOGrav signal as SIGWs, we utilize Bayesian analysis to obtain the model-independent reconstruction of the primordial curvature power spectrum that could explain the observed PTA signal. 
The model-independent reconstruction of the primordial curvature power spectrum is displayed through a free-spectrum framework that consists of 34 bins. 
The model-independent reconstructions are consistent with results obtained when using specific forms of the primordial curvature power spectrum, such as the lognormal and broken power-law forms, to fit the NANOGrav 15-year data set. 
Therefore, we can now just compare primordial curvature power spectra with the model-independent reconstruction to easily assess their compatibility with the PTA signals, without taking time to calculate the energy density of SIGWs.
This streamlines the assessment process and simplifies the task of determining whether a given primordial curvature power spectrum can adequately explain the observed PTA signals.

\section*{Acknowledgments}
ZY is supported by the National Natural Science Foundation of China under Grant No. 12205015 and the supporting fund for young researcher of Beijing Normal University under Grant No. 28719/310432102.
ZQY is supported by the National Natural Science Foundation of China under Grant No.~12305059.

\begin{thebibliography}{100}
	
	\bibitem{NANOGrav:2023gor}
	{\scshape NANOGrav} collaboration, \emph{{The NANOGrav 15 yr Data Set: Evidence
			for a Gravitational-wave Background}},
	\href{https://doi.org/10.3847/2041-8213/acdac6}{\emph{Astrophys. J. Lett.}
		{\bfseries 951} (2023) L8}
	[\href{https://arxiv.org/abs/2306.16213}{{\ttfamily 2306.16213}}].
	
	\bibitem{NANOGrav:2023hde}
	{\scshape NANOGrav} collaboration, \emph{{The NANOGrav 15 yr Data Set:
			Observations and Timing of 68 Millisecond Pulsars}},
	\href{https://doi.org/10.3847/2041-8213/acda9a}{\emph{Astrophys. J. Lett.}
		{\bfseries 951} (2023) L9}
	[\href{https://arxiv.org/abs/2306.16217}{{\ttfamily 2306.16217}}].
	
	\bibitem{Zic:2023gta}
	A.~Zic et~al., \emph{{The Parkes Pulsar Timing Array Third Data Release}},
	\href{https://arxiv.org/abs/2306.16230}{{\ttfamily 2306.16230}}.
	
	\bibitem{Reardon:2023gzh}
	D.J.~Reardon et~al., \emph{{Search for an Isotropic Gravitational-wave
			Background with the Parkes Pulsar Timing Array}},
	\href{https://doi.org/10.3847/2041-8213/acdd02}{\emph{Astrophys. J. Lett.}
		{\bfseries 951} (2023) L6}
	[\href{https://arxiv.org/abs/2306.16215}{{\ttfamily 2306.16215}}].
	
	\bibitem{Antoniadis:2023lym}
	{\scshape EPTA} collaboration, \emph{{The second data release from the European
			Pulsar Timing Array - I. The dataset and timing analysis}},
	\href{https://doi.org/10.1051/0004-6361/202346841}{\emph{Astron. Astrophys.}
		{\bfseries 678} (2023) A48}
	[\href{https://arxiv.org/abs/2306.16224}{{\ttfamily 2306.16224}}].
	
	\bibitem{Antoniadis:2023ott}
	{\scshape EPTA, InPTA:} collaboration, \emph{{The second data release from the
			European Pulsar Timing Array - III. Search for gravitational wave signals}},
	\href{https://doi.org/10.1051/0004-6361/202346844}{\emph{Astron. Astrophys.}
		{\bfseries 678} (2023) A50}
	[\href{https://arxiv.org/abs/2306.16214}{{\ttfamily 2306.16214}}].
	
	\bibitem{Xu:2023wog}
	H.~Xu et~al., \emph{{Searching for the Nano-Hertz Stochastic Gravitational Wave
			Background with the Chinese Pulsar Timing Array Data Release I}},
	\href{https://doi.org/10.1088/1674-4527/acdfa5}{\emph{Res. Astron.
			Astrophys.} {\bfseries 23} (2023) 075024}
	[\href{https://arxiv.org/abs/2306.16216}{{\ttfamily 2306.16216}}].
	
	\bibitem{NANOGrav:2023hvm}
	{\scshape NANOGrav} collaboration, \emph{{The NANOGrav 15 yr Data Set: Search
			for Signals from New Physics}},
	\href{https://doi.org/10.3847/2041-8213/acdc91}{\emph{Astrophys. J. Lett.}
		{\bfseries 951} (2023) L11}
	[\href{https://arxiv.org/abs/2306.16219}{{\ttfamily 2306.16219}}].
	
	\bibitem{Antoniadis:2023xlr}
	{\scshape EPTA} collaboration, \emph{{The second data release from the European
			Pulsar Timing Array: V. Implications for massive black holes, dark matter and
			the early Universe}},  \href{https://arxiv.org/abs/2306.16227}{{\ttfamily
			2306.16227}}.
	
	\bibitem{Yi:2023mbm}
	Z.~Yi, Q.~Gao, Y.~Gong, Y.~Wang and F.~Zhang, \emph{{Scalar induced
			gravitational waves in light of Pulsar Timing Array data}},
	\href{https://doi.org/10.1007/s11433-023-2266-1}{\emph{Sci. China Phys. Mech.
			Astron.} {\bfseries 66} (2023) 120404}
	[\href{https://arxiv.org/abs/2307.02467}{{\ttfamily 2307.02467}}].
	
	\bibitem{Franciolini:2023pbf}
	G.~Franciolini, A.~Iovino, Junior., V.~Vaskonen and H.~Veermae, \emph{{Recent
			Gravitational Wave Observation by Pulsar Timing Arrays and Primordial Black
			Holes: The Importance of Non-Gaussianities}},
	\href{https://doi.org/10.1103/PhysRevLett.131.201401}{\emph{Phys. Rev. Lett.}
		{\bfseries 131} (2023) 201401}
	[\href{https://arxiv.org/abs/2306.17149}{{\ttfamily 2306.17149}}].
	
	\bibitem{Liu:2023ymk}
	L.~Liu, Z.-C.~Chen and Q.-G.~Huang, \emph{{Implications for the non-Gaussianity
			of curvature perturbation from pulsar timing arrays}},
	\href{https://arxiv.org/abs/2307.01102}{{\ttfamily 2307.01102}}.
	
	\bibitem{Vagnozzi:2023lwo}
	S.~Vagnozzi, \emph{{Inflationary interpretation of the stochastic gravitational
			wave background signal detected by pulsar timing array experiments}},
	\href{https://doi.org/10.1016/j.jheap.2023.07.001}{\emph{JHEAp} {\bfseries
			39} (2023) 81} [\href{https://arxiv.org/abs/2306.16912}{{\ttfamily
			2306.16912}}].
	
	\bibitem{Cai:2023dls}
	Y.-F.~Cai, X.-C.~He, X.-H.~Ma, S.-F.~Yan and G.-W.~Yuan, \emph{{Limits on
			scalar-induced gravitational waves from the stochastic background by pulsar
			timing array observations}},
	\href{https://arxiv.org/abs/2306.17822}{{\ttfamily 2306.17822}}.
	
	\bibitem{Bi:2023tib}
	Y.-C.~Bi, Y.-M.~Wu, Z.-C.~Chen and Q.-G.~Huang, \emph{{Implications for the
			supermassive black hole binaries from the NANOGrav 15-year data set}},
	\href{https://doi.org/10.1007/s11433-023-2252-4}{\emph{Sci. China Phys. Mech.
			Astron.} {\bfseries 66} (2023) 120402}
	[\href{https://arxiv.org/abs/2307.00722}{{\ttfamily 2307.00722}}].
	
	\bibitem{Wu:2023hsa}
	Y.-M.~Wu, Z.-C.~Chen and Q.-G.~Huang, \emph{{Cosmological Interpretation for
			the Stochastic Signal in Pulsar Timing Arrays}},
	\href{https://arxiv.org/abs/2307.03141}{{\ttfamily 2307.03141}}.
	
	\bibitem{Franciolini:2023wjm}
	G.~Franciolini, D.~Racco and F.~Rompineve, \emph{{Footprints of the QCD
			Crossover on Cosmological Gravitational Waves at Pulsar Timing Arrays}},
	\href{https://arxiv.org/abs/2306.17136}{{\ttfamily 2306.17136}}.
	
	\bibitem{You:2023rmn}
	Z.-Q.~You, Z.~Yi and Y.~Wu, \emph{{Constraints on primordial curvature power
			spectrum with pulsar timing arrays}},
	\href{https://doi.org/10.1088/1475-7516/2023/11/065}{\emph{JCAP} {\bfseries
			11} (2023) 065} [\href{https://arxiv.org/abs/2307.04419}{{\ttfamily
			2307.04419}}].
	
	\bibitem{Jin:2023wri}
	J.-H.~Jin, Z.-C.~Chen, Z.~Yi, Z.-Q.~You, L.~Liu and Y.~Wu, \emph{{Confronting
			sound speed resonance with pulsar timing arrays}},
	\href{https://doi.org/10.1088/1475-7516/2023/09/016}{\emph{JCAP} {\bfseries
			09} (2023) 016} [\href{https://arxiv.org/abs/2307.08687}{{\ttfamily
			2307.08687}}].
	
	\bibitem{Liu:2023pau}
	L.~Liu, Z.-C.~Chen and Q.-G.~Huang, \emph{{Probing the equation of state of the
			early Universe with pulsar timing arrays}},
	\href{https://doi.org/10.1088/1475-7516/2023/11/071}{\emph{JCAP} {\bfseries
			11} (2023) 071} [\href{https://arxiv.org/abs/2307.14911}{{\ttfamily
			2307.14911}}].
	
	\bibitem{An:2023jxf}
	H.~An, B.~Su, H.~Tai, L.-T.~Wang and C.~Yang, \emph{{Phase transition during
			inflation and the gravitational wave signal at pulsar timing arrays}},
	\href{https://arxiv.org/abs/2308.00070}{{\ttfamily 2308.00070}}.
	
	\bibitem{Zhang:2023nrs}
	Z.~Zhang, C.~Cai, Y.-H.~Su, S.~Wang, Z.-H.~Yu and H.-H.~Zhang,
	\emph{{Nano-Hertz gravitational waves from collapsing domain walls associated
			with freeze-in dark matter in light of pulsar timing array observations}},
	\href{https://doi.org/10.1103/PhysRevD.108.095037}{\emph{Phys. Rev. D}
		{\bfseries 108} (2023) 095037}
	[\href{https://arxiv.org/abs/2307.11495}{{\ttfamily 2307.11495}}].
	
	\bibitem{Das:2023nmm}
	B.~Das, N.~Jaman and M.~Sami, \emph{{Gravitational wave background from
			quintessential inflation and NANOGrav data}},
	\href{https://doi.org/10.1103/PhysRevD.108.103510}{\emph{Phys. Rev. D}
		{\bfseries 108} (2023) 103510}
	[\href{https://arxiv.org/abs/2307.12913}{{\ttfamily 2307.12913}}].
	
	\bibitem{Balaji:2023ehk}
	S.~Balaji, G.~Dom\`enech and G.~Franciolini, \emph{{Scalar-induced
			gravitational wave interpretation of PTA data: the role of scalar fluctuation
			propagation speed}},
	\href{https://doi.org/10.1088/1475-7516/2023/10/041}{\emph{JCAP} {\bfseries
			10} (2023) 041} [\href{https://arxiv.org/abs/2307.08552}{{\ttfamily
			2307.08552}}].
	
	\bibitem{Du:2023qvj}
	X.K.~Du, M.X.~Huang, F.~Wang and Y.K.~Zhang, \emph{{Did the nHZ Gravitational
			Waves Signatures Observed By NANOGrav Indicate Multiple Sector SUSY
			Breaking?}},  \href{https://arxiv.org/abs/2307.02938}{{\ttfamily
			2307.02938}}.
	
	\bibitem{Oikonomou:2023qfz}
	V.K.~Oikonomou, \emph{{Flat energy spectrum of primordial gravitational waves
			versus peaks and the NANOGrav 2023 observation}},
	\href{https://doi.org/10.1103/PhysRevD.108.043516}{\emph{Phys. Rev. D}
		{\bfseries 108} (2023) 043516}
	[\href{https://arxiv.org/abs/2306.17351}{{\ttfamily 2306.17351}}].
	
	\bibitem{Yi:2023npi}
	Z.~Yi, Z.-Q.~You, Y.~Wu, Z.-C.~Chen and L.~Liu, \emph{{Exploring the NANOGrav
			Signal and Planet-mass Primordial Black Holes through Higgs Inflation}},
	\href{https://arxiv.org/abs/2308.14688}{{\ttfamily 2308.14688}}.
	
	\bibitem{Chen:2023zkb}
	Z.-C.~Chen, Q.-G.~Huang, C.~Liu, L.~Liu, X.-J.~Liu, Y.~Wu et~al.,
	\emph{{Prospects for Taiji to detect a gravitational-wave background from
			cosmic strings}},  \href{https://arxiv.org/abs/2310.00411}{{\ttfamily
			2310.00411}}.
	
	\bibitem{Liu:2023hpw}
	L.~Liu, Y.~Wu and Z.-C.~Chen, \emph{{Simultaneously probing the sound speed and
			equation of state of the early Universe with pulsar timing arrays}},
	\href{https://arxiv.org/abs/2310.16500}{{\ttfamily 2310.16500}}.
	
	\bibitem{Zhu:2018lif}
	X.-J.~Zhu, W.~Cui and E.~Thrane, \emph{{The minimum and maximum
			gravitational-wave background from supermassive binary black holes}},
	\href{https://doi.org/10.1093/mnras/sty2849}{\emph{Mon. Not. Roy. Astron.
			Soc.} {\bfseries 482} (2019) 2588}
	[\href{https://arxiv.org/abs/1806.02346}{{\ttfamily 1806.02346}}].
	
	\bibitem{Li:2019vlb}
	J.~Li, Z.-C.~Chen and Q.-G.~Huang, \emph{{Measuring the tilt of primordial
			gravitational-wave power spectrum from observations}},
	\href{https://doi.org/10.1007/s11433-019-9605-5}{\emph{Sci. China Phys. Mech.
			Astron.} {\bfseries 62} (2019) 110421}
	[\href{https://arxiv.org/abs/1907.09794}{{\ttfamily 1907.09794}}].
	
	\bibitem{Chen:2021wdo}
	Z.-C.~Chen, C.~Yuan and Q.-G.~Huang, \emph{{Non-tensorial gravitational wave
			background in NANOGrav 12.5-year data set}},
	\href{https://doi.org/10.1007/s11433-021-1797-y}{\emph{Sci. China Phys. Mech.
			Astron.} {\bfseries 64} (2021) 120412}
	[\href{https://arxiv.org/abs/2101.06869}{{\ttfamily 2101.06869}}].
	
	\bibitem{Wu:2021kmd}
	Y.-M.~Wu, Z.-C.~Chen and Q.-G.~Huang, \emph{{Constraining the Polarization of
			Gravitational Waves with the Parkes Pulsar Timing Array Second Data
			Release}}, \href{https://doi.org/10.3847/1538-4357/ac35cc}{\emph{Astrophys.
			J.} {\bfseries 925} (2022) 37}
	[\href{https://arxiv.org/abs/2108.10518}{{\ttfamily 2108.10518}}].
	
	\bibitem{Chen:2021ncc}
	Z.-C.~Chen, Y.-M.~Wu and Q.-G.~Huang, \emph{{Searching for isotropic stochastic
			gravitational-wave background in the international pulsar timing array second
			data release}}, \href{https://doi.org/10.1088/1572-9494/ac7cdf}{\emph{Commun.
			Theor. Phys.} {\bfseries 74} (2022) 105402}
	[\href{https://arxiv.org/abs/2109.00296}{{\ttfamily 2109.00296}}].
	
	\bibitem{Chen:2022azo}
	Z.-C.~Chen, Y.-M.~Wu and Q.-G.~Huang, \emph{{Search for the Gravitational-wave
			Background from Cosmic Strings with the Parkes Pulsar Timing Array Second
			Data Release}},
	\href{https://doi.org/10.3847/1538-4357/ac86cb}{\emph{Astrophys. J.}
		{\bfseries 936} (2022) 20}
	[\href{https://arxiv.org/abs/2205.07194}{{\ttfamily 2205.07194}}].
	
	\bibitem{PPTA:2022eul}
	{\scshape PPTA} collaboration, \emph{{Constraining ultralight vector dark
			matter with the Parkes Pulsar Timing Array second data release}},
	\href{https://doi.org/10.1103/PhysRevD.106.L081101}{\emph{Phys. Rev. D}
		{\bfseries 106} (2022) L081101}
	[\href{https://arxiv.org/abs/2210.03880}{{\ttfamily 2210.03880}}].
	
	\bibitem{IPTA:2023ero}
	{\scshape IPTA} collaboration, \emph{{Searching for continuous Gravitational
			Waves in the second data release of the International Pulsar Timing Array}},
	\href{https://doi.org/10.1093/mnras/stad812}{\emph{Mon. Not. Roy. Astron.
			Soc.} {\bfseries 521} (2023) 5077}
	[\href{https://arxiv.org/abs/2303.10767}{{\ttfamily 2303.10767}}].
	
	\bibitem{Wu:2023pbt}
	Y.-M.~Wu, Z.-C.~Chen and Q.-G.~Huang, \emph{{Search for stochastic
			gravitational-wave background from massive gravity in the NANOGrav 12.5-year
			dataset}}, \href{https://doi.org/10.1103/PhysRevD.107.042003}{\emph{Phys.
			Rev. D} {\bfseries 107} (2023) 042003}
	[\href{https://arxiv.org/abs/2302.00229}{{\ttfamily 2302.00229}}].
	
	\bibitem{Wu:2023dnp}
	Y.-M.~Wu, Z.-C.~Chen and Q.-G.~Huang, \emph{{Pulsar timing residual induced by
			ultralight tensor dark matter}},
	\href{https://doi.org/10.1088/1475-7516/2023/09/021}{\emph{JCAP} {\bfseries
			09} (2023) 021} [\href{https://arxiv.org/abs/2305.08091}{{\ttfamily
			2305.08091}}].
	
	\bibitem{InternationalPulsarTimingArray:2023mzf}
	{\scshape International Pulsar Timing Array} collaboration, \emph{{Comparing
			recent PTA results on the nanohertz stochastic gravitational wave
			background}},  \href{https://arxiv.org/abs/2309.00693}{{\ttfamily
			2309.00693}}.
	
	\bibitem{Wu:2023rib}
	Y.-M.~Wu, Z.-C.~Chen, Y.-C.~Bi and Q.-G.~Huang, \emph{{Constraining the
			Graviton Mass with the NANOGrav 15-Year Data Set}},
	\href{https://arxiv.org/abs/2310.07469}{{\ttfamily 2310.07469}}.
	
	\bibitem{Bi:2023ewq}
	Y.-C.~Bi, Y.-M.~Wu, Z.-C.~Chen and Q.-G.~Huang, \emph{{Constraints on the
			velocity of gravitational waves from NANOGrav 15-year data set}},
	\href{https://arxiv.org/abs/2310.08366}{{\ttfamily 2310.08366}}.
	
	\bibitem{Chen:2023uiz}
	Z.-C.~Chen, Y.-M.~Wu, Y.-C.~Bi and Q.-G.~Huang, \emph{{Search for Non-Tensorial
			Gravitational-Wave Backgrounds in the NANOGrav 15-Year Data Set}},
	\href{https://arxiv.org/abs/2310.11238}{{\ttfamily 2310.11238}}.
	
	\bibitem{Hawking:1971ei}
	S.~Hawking, \emph{{Gravitationally collapsed objects of very low mass}},
	\href{https://doi.org/10.1093/mnras/152.1.75}{\emph{Mon. Not. Roy. Astron.
			Soc.} {\bfseries 152} (1971) 75}.
	
	\bibitem{Carr:1974nx}
	B.J.~Carr and S.W.~Hawking, \emph{{Black holes in the early Universe}},
	\href{https://doi.org/10.1093/mnras/168.2.399}{\emph{Mon. Not. Roy. Astron.
			Soc.} {\bfseries 168} (1974) 399}.
	
	\bibitem{Ananda:2006af}
	K.N.~Ananda, C.~Clarkson and D.~Wands, \emph{{The Cosmological gravitational
			wave background from primordial density perturbations}},
	\href{https://doi.org/10.1103/PhysRevD.75.123518}{\emph{Phys. Rev. D}
		{\bfseries 75} (2007) 123518}
	[\href{https://arxiv.org/abs/gr-qc/0612013}{{\ttfamily gr-qc/0612013}}].
	
	\bibitem{Baumann:2007zm}
	D.~Baumann, P.J.~Steinhardt, K.~Takahashi and K.~Ichiki, \emph{{Gravitational
			Wave Spectrum Induced by Primordial Scalar Perturbations}},
	\href{https://doi.org/10.1103/PhysRevD.76.084019}{\emph{Phys. Rev. D}
		{\bfseries 76} (2007) 084019}
	[\href{https://arxiv.org/abs/hep-th/0703290}{{\ttfamily hep-th/0703290}}].
	
	\bibitem{Saito:2008jc}
	R.~Saito and J.~Yokoyama, \emph{{Gravitational wave background as a probe of
			the primordial black hole abundance}},
	\href{https://doi.org/10.1103/PhysRevLett.102.161101}{\emph{Phys. Rev. Lett.}
		{\bfseries 102} (2009) 161101}
	[\href{https://arxiv.org/abs/0812.4339}{{\ttfamily 0812.4339}}].
	
	\bibitem{Alabidi:2012ex}
	L.~Alabidi, K.~Kohri, M.~Sasaki and Y.~Sendouda, \emph{{Observable Spectra of
			Induced Gravitational Waves from Inflation}},
	\href{https://doi.org/10.1088/1475-7516/2012/09/017}{\emph{JCAP} {\bfseries
			09} (2012) 017} [\href{https://arxiv.org/abs/1203.4663}{{\ttfamily
			1203.4663}}].
	
	\bibitem{Sasaki:2018dmp}
	M.~Sasaki, T.~Suyama, T.~Tanaka and S.~Yokoyama, \emph{{Primordial black
			holes\textemdash{}perspectives in gravitational wave astronomy}},
	\href{https://doi.org/10.1088/1361-6382/aaa7b4}{\emph{Class. Quant. Grav.}
		{\bfseries 35} (2018) 063001}
	[\href{https://arxiv.org/abs/1801.05235}{{\ttfamily 1801.05235}}].
	
	\bibitem{Nakama:2016gzw}
	T.~Nakama, J.~Silk and M.~Kamionkowski, \emph{{Stochastic gravitational waves
			associated with the formation of primordial black holes}},
	\href{https://doi.org/10.1103/PhysRevD.95.043511}{\emph{Phys. Rev. D}
		{\bfseries 95} (2017) 043511}
	[\href{https://arxiv.org/abs/1612.06264}{{\ttfamily 1612.06264}}].
	
	\bibitem{Kohri:2018awv}
	K.~Kohri and T.~Terada, \emph{{Semianalytic calculation of gravitational wave
			spectrum nonlinearly induced from primordial curvature perturbations}},
	\href{https://doi.org/10.1103/PhysRevD.97.123532}{\emph{Phys. Rev. D}
		{\bfseries 97} (2018) 123532}
	[\href{https://arxiv.org/abs/1804.08577}{{\ttfamily 1804.08577}}].
	
	\bibitem{Di:2017ndc}
	H.~Di and Y.~Gong, \emph{{Primordial black holes and second order gravitational
			waves from ultra-slow-roll inflation}},
	\href{https://doi.org/10.1088/1475-7516/2018/07/007}{\emph{JCAP} {\bfseries
			07} (2018) 007} [\href{https://arxiv.org/abs/1707.09578}{{\ttfamily
			1707.09578}}].
	
	\bibitem{Cheng:2018yyr}
	S.-L.~Cheng, W.~Lee and K.-W.~Ng, \emph{{Primordial black holes and associated
			gravitational waves in axion monodromy inflation}},
	\href{https://doi.org/10.1088/1475-7516/2018/07/001}{\emph{JCAP} {\bfseries
			07} (2018) 001} [\href{https://arxiv.org/abs/1801.09050}{{\ttfamily
			1801.09050}}].
	
	\bibitem{Cai:2019amo}
	R.-G.~Cai, S.~Pi, S.-J.~Wang and X.-Y.~Yang, \emph{{Resonant multiple peaks in
			the induced gravitational waves}},
	\href{https://doi.org/10.1088/1475-7516/2019/05/013}{\emph{JCAP} {\bfseries
			05} (2019) 013} [\href{https://arxiv.org/abs/1901.10152}{{\ttfamily
			1901.10152}}].
	
	\bibitem{Cai:2018dig}
	R.-g.~Cai, S.~Pi and M.~Sasaki, \emph{{Gravitational Waves Induced by
			non-Gaussian Scalar Perturbations}},
	\href{https://doi.org/10.1103/PhysRevLett.122.201101}{\emph{Phys. Rev. Lett.}
		{\bfseries 122} (2019) 201101}
	[\href{https://arxiv.org/abs/1810.11000}{{\ttfamily 1810.11000}}].
	
	\bibitem{Cai:2019elf}
	R.-G.~Cai, S.~Pi, S.-J.~Wang and X.-Y.~Yang, \emph{{Pulsar Timing Array
			Constraints on the Induced Gravitational Waves}},
	\href{https://doi.org/10.1088/1475-7516/2019/10/059}{\emph{JCAP} {\bfseries
			10} (2019) 059} [\href{https://arxiv.org/abs/1907.06372}{{\ttfamily
			1907.06372}}].
	
	\bibitem{Cai:2019bmk}
	R.-G.~Cai, Z.-K.~Guo, J.~Liu, L.~Liu and X.-Y.~Yang, \emph{{Primordial black
			holes and gravitational waves from parametric amplification of curvature
			perturbations}},
	\href{https://doi.org/10.1088/1475-7516/2020/06/013}{\emph{JCAP} {\bfseries
			06} (2020) 013} [\href{https://arxiv.org/abs/1912.10437}{{\ttfamily
			1912.10437}}].
	
	\bibitem{Wu:2020drm}
	Y.~Wu, \emph{{Merger history of primordial black-hole binaries}},
	\href{https://doi.org/10.1103/PhysRevD.101.083008}{\emph{Phys. Rev. D}
		{\bfseries 101} (2020) 083008}
	[\href{https://arxiv.org/abs/2001.03833}{{\ttfamily 2001.03833}}].
	
	\bibitem{Cai:2020fnq}
	R.-G.~Cai, Y.-C.~Ding, X.-Y.~Yang and Y.-F.~Zhou, \emph{{Constraints on a mixed
			model of dark matter particles and primordial black holes from the galactic
			511 keV line}},
	\href{https://doi.org/10.1088/1475-7516/2021/03/057}{\emph{JCAP} {\bfseries
			03} (2021) 057} [\href{https://arxiv.org/abs/2007.11804}{{\ttfamily
			2007.11804}}].
	
	\bibitem{Pi:2020otn}
	S.~Pi and M.~Sasaki, \emph{{Gravitational Waves Induced by Scalar Perturbations
			with a Lognormal Peak}},
	\href{https://doi.org/10.1088/1475-7516/2020/09/037}{\emph{JCAP} {\bfseries
			09} (2020) 037} [\href{https://arxiv.org/abs/2005.12306}{{\ttfamily
			2005.12306}}].
	
	\bibitem{Domenech:2020kqm}
	G.~Dom\`enech, S.~Pi and M.~Sasaki, \emph{{Induced gravitational waves as a
			probe of thermal history of the universe}},
	\href{https://doi.org/10.1088/1475-7516/2020/08/017}{\emph{JCAP} {\bfseries
			08} (2020) 017} [\href{https://arxiv.org/abs/2005.12314}{{\ttfamily
			2005.12314}}].
	
	\bibitem{Liu:2018ess}
	L.~Liu, Z.-K.~Guo and R.-G.~Cai, \emph{{Effects of the surrounding primordial
			black holes on the merger rate of primordial black hole binaries}},
	\href{https://doi.org/10.1103/PhysRevD.99.063523}{\emph{Phys. Rev. D}
		{\bfseries 99} (2019) 063523}
	[\href{https://arxiv.org/abs/1812.05376}{{\ttfamily 1812.05376}}].
	
	\bibitem{Liu:2019rnx}
	L.~Liu, Z.-K.~Guo and R.-G.~Cai, \emph{{Effects of the merger history on the
			merger rate density of primordial black hole binaries}},
	\href{https://doi.org/10.1140/epjc/s10052-019-7227-0}{\emph{Eur. Phys. J. C}
		{\bfseries 79} (2019) 717}
	[\href{https://arxiv.org/abs/1901.07672}{{\ttfamily 1901.07672}}].
	
	\bibitem{Liu:2020cds}
	L.~Liu, Z.-K.~Guo, R.-G.~Cai and S.P.~Kim, \emph{{Merger rate distribution of
			primordial black hole binaries with electric charges}},
	\href{https://doi.org/10.1103/PhysRevD.102.043508}{\emph{Phys. Rev. D}
		{\bfseries 102} (2020) 043508}
	[\href{https://arxiv.org/abs/2001.02984}{{\ttfamily 2001.02984}}].
	
	\bibitem{Liu:2021jnw}
	L.~Liu, X.-Y.~Yang, Z.-K.~Guo and R.-G.~Cai, \emph{{Testing primordial black
			hole and measuring the Hubble constant with multiband gravitational-wave
			observations}},
	\href{https://doi.org/10.1088/1475-7516/2023/01/006}{\emph{JCAP} {\bfseries
			01} (2023) 006} [\href{https://arxiv.org/abs/2112.05473}{{\ttfamily
			2112.05473}}].
	
	\bibitem{Liu:2022iuf}
	L.~Liu, Z.-Q.~You, Y.~Wu and Z.-C.~Chen, \emph{{Constraining the merger history
			of primordial-black-hole binaries from GWTC-3}},
	\href{https://doi.org/10.1103/PhysRevD.107.063035}{\emph{Phys. Rev. D}
		{\bfseries 107} (2023) 063035}
	[\href{https://arxiv.org/abs/2210.16094}{{\ttfamily 2210.16094}}].
	
	\bibitem{Chen:2019xse}
	Z.-C.~Chen, C.~Yuan and Q.-G.~Huang, \emph{{Pulsar Timing Array Constraints on
			Primordial Black Holes with NANOGrav 11-Year Dataset}},
	\href{https://doi.org/10.1103/PhysRevLett.124.251101}{\emph{Phys. Rev. Lett.}
		{\bfseries 124} (2020) 251101}
	[\href{https://arxiv.org/abs/1910.12239}{{\ttfamily 1910.12239}}].
	
	\bibitem{Yuan:2019fwv}
	C.~Yuan, Z.-C.~Chen and Q.-G.~Huang, \emph{{Scalar induced gravitational waves
			in different gauges}},
	\href{https://doi.org/10.1103/PhysRevD.101.063018}{\emph{Phys. Rev. D}
		{\bfseries 101} (2020) 063018}
	[\href{https://arxiv.org/abs/1912.00885}{{\ttfamily 1912.00885}}].
	
	\bibitem{Yuan:2019wwo}
	C.~Yuan, Z.-C.~Chen and Q.-G.~Huang, \emph{{Log-dependent slope of scalar
			induced gravitational waves in the infrared regions}},
	\href{https://doi.org/10.1103/PhysRevD.101.043019}{\emph{Phys. Rev. D}
		{\bfseries 101} (2020) 043019}
	[\href{https://arxiv.org/abs/1910.09099}{{\ttfamily 1910.09099}}].
	
	\bibitem{Yuan:2019udt}
	C.~Yuan, Z.-C.~Chen and Q.-G.~Huang, \emph{{Probing
			primordial\textendash{}black-hole dark matter with scalar induced
			gravitational waves}},
	\href{https://doi.org/10.1103/PhysRevD.100.081301}{\emph{Phys. Rev. D}
		{\bfseries 100} (2019) 081301}
	[\href{https://arxiv.org/abs/1906.11549}{{\ttfamily 1906.11549}}].
	
	\bibitem{Carr:2020gox}
	B.~Carr, K.~Kohri, Y.~Sendouda and J.~Yokoyama, \emph{{Constraints on
			primordial black holes}},
	\href{https://doi.org/10.1088/1361-6633/ac1e31}{\emph{Rept. Prog. Phys.}
		{\bfseries 84} (2021) 116902}
	[\href{https://arxiv.org/abs/2002.12778}{{\ttfamily 2002.12778}}].
	
	\bibitem{Liu:2020vsy}
	L.~Liu, O.~Christiansen, Z.-K.~Guo, R.-G.~Cai and S.P.~Kim,
	\emph{{Gravitational and electromagnetic radiation from binary black holes
			with electric and magnetic charges: Circular orbits on a cone}},
	\href{https://doi.org/10.1103/PhysRevD.102.103520}{\emph{Phys. Rev. D}
		{\bfseries 102} (2020) 103520}
	[\href{https://arxiv.org/abs/2008.02326}{{\ttfamily 2008.02326}}].
	
	\bibitem{Liu:2020bag}
	L.~Liu, O.~Christiansen, W.-H.~Ruan, Z.-K.~Guo, R.-G.~Cai and S.P.~Kim,
	\emph{{Gravitational and electromagnetic radiation from binary black holes
			with electric and magnetic charges: elliptical orbits on a cone}},
	\href{https://doi.org/10.1140/epjc/s10052-021-09849-4}{\emph{Eur. Phys. J. C}
		{\bfseries 81} (2021) 1048}
	[\href{https://arxiv.org/abs/2011.13586}{{\ttfamily 2011.13586}}].
	
	\bibitem{Papanikolaou:2021uhe}
	T.~Papanikolaou, C.~Tzerefos, S.~Basilakos and E.N.~Saridakis, \emph{{Scalar
			induced gravitational waves from primordial black hole Poisson fluctuations
			in f(R) gravity}},
	\href{https://doi.org/10.1088/1475-7516/2022/10/013}{\emph{JCAP} {\bfseries
			10} (2022) 013} [\href{https://arxiv.org/abs/2112.15059}{{\ttfamily
			2112.15059}}].
	
	\bibitem{Papanikolaou:2022hkg}
	T.~Papanikolaou, C.~Tzerefos, S.~Basilakos and E.N.~Saridakis, \emph{{No
			constraints for f(T) gravity from gravitational waves induced from primordial
			black hole fluctuations}},
	\href{https://doi.org/10.1140/epjc/s10052-022-11157-4}{\emph{Eur. Phys. J. C}
		{\bfseries 83} (2023) 31} [\href{https://arxiv.org/abs/2205.06094}{{\ttfamily
			2205.06094}}].
	
	\bibitem{Chakraborty:2022mwu}
	A.~Chakraborty, P.K.~Chanda, K.L.~Pandey and S.~Das, \emph{{Formation and
			Abundance of Late-forming Primordial Black Holes as Dark Matter}},
	\href{https://doi.org/10.3847/1538-4357/ac6ddd}{\emph{Astrophys. J.}
		{\bfseries 932} (2022) 119}
	[\href{https://arxiv.org/abs/2204.09628}{{\ttfamily 2204.09628}}].
	
	\bibitem{Chen:2018czv}
	Z.-C.~Chen and Q.-G.~Huang, \emph{{Merger Rate Distribution of
			Primordial-Black-Hole Binaries}},
	\href{https://doi.org/10.3847/1538-4357/aad6e2}{\emph{Astrophys. J.}
		{\bfseries 864} (2018) 61}
	[\href{https://arxiv.org/abs/1801.10327}{{\ttfamily 1801.10327}}].
	
	\bibitem{Chen:2018rzo}
	Z.-C.~Chen, F.~Huang and Q.-G.~Huang, \emph{{Stochastic Gravitational-wave
			Background from Binary Black Holes and Binary Neutron Stars and Implications
			for LISA}}, \href{https://doi.org/10.3847/1538-4357/aaf581}{\emph{Astrophys.
			J.} {\bfseries 871} (2019) 97}
	[\href{https://arxiv.org/abs/1809.10360}{{\ttfamily 1809.10360}}].
	
	\bibitem{Chen:2019irf}
	Z.-C.~Chen and Q.-G.~Huang, \emph{{Distinguishing Primordial Black Holes from
			Astrophysical Black Holes by Einstein Telescope and Cosmic Explorer}},
	\href{https://doi.org/10.1088/1475-7516/2020/08/039}{\emph{JCAP} {\bfseries
			08} (2020) 039} [\href{https://arxiv.org/abs/1904.02396}{{\ttfamily
			1904.02396}}].
	
	\bibitem{Chen:2021nxo}
	Z.-C.~Chen, C.~Yuan and Q.-G.~Huang, \emph{{Confronting the primordial black
			hole scenario with the gravitational-wave events detected by LIGO-Virgo}},
	\href{https://doi.org/10.1016/j.physletb.2022.137040}{\emph{Phys. Lett. B}
		{\bfseries 829} (2022) 137040}
	[\href{https://arxiv.org/abs/2108.11740}{{\ttfamily 2108.11740}}].
	
	\bibitem{Chen:2022fda}
	Z.-C.~Chen, S.-S.~Du, Q.-G.~Huang and Z.-Q.~You, \emph{{Constraints on
			primordial-black-hole population and cosmic expansion history from GWTC-3}},
	\href{https://doi.org/10.1088/1475-7516/2023/03/024}{\emph{JCAP} {\bfseries
			03} (2023) 024} [\href{https://arxiv.org/abs/2205.11278}{{\ttfamily
			2205.11278}}].
	
	\bibitem{Liu:2022wtq}
	L.~Liu and S.P.~Kim, \emph{{Merger rate of charged black holes from the
			two-body dynamical capture}},
	\href{https://doi.org/10.1088/1475-7516/2022/03/059}{\emph{JCAP} {\bfseries
			03} (2022) 059} [\href{https://arxiv.org/abs/2201.02581}{{\ttfamily
			2201.02581}}].
	
	\bibitem{Zheng:2022wqo}
	L.-M.~Zheng, Z.~Li, Z.-C.~Chen, H.~Zhou and Z.-H.~Zhu, \emph{{Towards a
			reliable reconstruction of the power spectrum of primordial curvature
			perturbation on small scales from GWTC-3}},
	\href{https://doi.org/10.1016/j.physletb.2023.137720}{\emph{Phys. Lett. B}
		{\bfseries 838} (2023) 137720}
	[\href{https://arxiv.org/abs/2212.05516}{{\ttfamily 2212.05516}}].
	
	\bibitem{Chen:2022qvg}
	Z.-C.~Chen, S.P.~Kim and L.~Liu, \emph{{Gravitational and electromagnetic
			radiation from binary black holes with electric and magnetic charges:
			hyperbolic orbits on a cone}},
	\href{https://doi.org/10.1088/1572-9494/acce98}{\emph{Commun. Theor. Phys.}
		{\bfseries 75} (2023) 065401}
	[\href{https://arxiv.org/abs/2210.15564}{{\ttfamily 2210.15564}}].
	
	\bibitem{Meng:2022low}
	D.-S.~Meng, C.~Yuan and Q.-G.~Huang, \emph{{Primordial black holes generated by
			the non-minimal spectator field}},
	\href{https://doi.org/10.1007/s11433-022-2095-5}{\emph{Sci. China Phys. Mech.
			Astron.} {\bfseries 66} (2023) 280411}
	[\href{https://arxiv.org/abs/2212.03577}{{\ttfamily 2212.03577}}].
	
	\bibitem{Garcia-Saenz:2023zue}
	S.~Garcia-Saenz, Y.~Lu and Z.~Shuai, \emph{{Scalar-Induced Gravitational Waves
			from Ghost Inflation and Parity Violation}},
	\href{https://arxiv.org/abs/2306.09052}{{\ttfamily 2306.09052}}.
	
	\bibitem{Planck:2018jri}
	{\scshape Planck} collaboration, \emph{{Planck 2018 results. X. Constraints on
			inflation}}, \href{https://doi.org/10.1051/0004-6361/201833887}{\emph{Astron.
			Astrophys.} {\bfseries 641} (2020) A10}
	[\href{https://arxiv.org/abs/1807.06211}{{\ttfamily 1807.06211}}].
	
	\bibitem{Martin:2012pe}
	J.~Martin, H.~Motohashi and T.~Suyama, \emph{{Ultra Slow-Roll Inflation and the
			non-Gaussianity Consistency Relation}},
	\href{https://doi.org/10.1103/PhysRevD.87.023514}{\emph{Phys. Rev. D}
		{\bfseries 87} (2013) 023514}
	[\href{https://arxiv.org/abs/1211.0083}{{\ttfamily 1211.0083}}].
	
	\bibitem{Motohashi:2014ppa}
	H.~Motohashi, A.A.~Starobinsky and J.~Yokoyama, \emph{{Inflation with a
			constant rate of roll}},
	\href{https://doi.org/10.1088/1475-7516/2015/09/018}{\emph{JCAP} {\bfseries
			09} (2015) 018} [\href{https://arxiv.org/abs/1411.5021}{{\ttfamily
			1411.5021}}].
	
	\bibitem{Yi:2017mxs}
	Z.~Yi and Y.~Gong, \emph{{On the constant-roll inflation}},
	\href{https://doi.org/10.1088/1475-7516/2018/03/052}{\emph{JCAP} {\bfseries
			03} (2018) 052} [\href{https://arxiv.org/abs/1712.07478}{{\ttfamily
			1712.07478}}].
	
	\bibitem{Garcia-Bellido:2017mdw}
	J.~Garcia-Bellido and E.~Ruiz~Morales, \emph{{Primordial black holes from
			single field models of inflation}},
	\href{https://doi.org/10.1016/j.dark.2017.09.007}{\emph{Phys. Dark Univ.}
		{\bfseries 18} (2017) 47} [\href{https://arxiv.org/abs/1702.03901}{{\ttfamily
			1702.03901}}].
	
	\bibitem{Germani:2017bcs}
	C.~Germani and T.~Prokopec, \emph{{On primordial black holes from an inflection
			point}}, \href{https://doi.org/10.1016/j.dark.2017.09.001}{\emph{Phys. Dark
			Univ.} {\bfseries 18} (2017) 6}
	[\href{https://arxiv.org/abs/1706.04226}{{\ttfamily 1706.04226}}].
	
	\bibitem{Motohashi:2017kbs}
	H.~Motohashi and W.~Hu, \emph{{Primordial Black Holes and Slow-Roll
			Violation}}, \href{https://doi.org/10.1103/PhysRevD.96.063503}{\emph{Phys.
			Rev. D} {\bfseries 96} (2017) 063503}
	[\href{https://arxiv.org/abs/1706.06784}{{\ttfamily 1706.06784}}].
	
	\bibitem{Ezquiaga:2017fvi}
	J.M.~Ezquiaga, J.~Garcia-Bellido and E.~Ruiz~Morales, \emph{{Primordial Black
			Hole production in Critical Higgs Inflation}},
	\href{https://doi.org/10.1016/j.physletb.2017.11.039}{\emph{Phys. Lett. B}
		{\bfseries 776} (2018) 345}
	[\href{https://arxiv.org/abs/1705.04861}{{\ttfamily 1705.04861}}].
	
	\bibitem{Gong:2017qlj}
	H.~Di and Y.~Gong, \emph{{Primordial black holes and second order gravitational
			waves from ultra-slow-roll inflation}},
	\href{https://doi.org/10.1088/1475-7516/2018/07/007}{\emph{JCAP} {\bfseries
			07} (2018) 007} [\href{https://arxiv.org/abs/1707.09578}{{\ttfamily
			1707.09578}}].
	
	\bibitem{Ballesteros:2018wlw}
	G.~Ballesteros, J.~Beltran~Jimenez and M.~Pieroni, \emph{{Black hole formation
			from a general quadratic action for inflationary primordial fluctuations}},
	\href{https://doi.org/10.1088/1475-7516/2019/06/016}{\emph{JCAP} {\bfseries
			06} (2019) 016} [\href{https://arxiv.org/abs/1811.03065}{{\ttfamily
			1811.03065}}].
	
	\bibitem{Dalianis:2018frf}
	I.~Dalianis, A.~Kehagias and G.~Tringas, \emph{{Primordial black holes from
			\ensuremath{\alpha}-attractors}},
	\href{https://doi.org/10.1088/1475-7516/2019/01/037}{\emph{JCAP} {\bfseries
			01} (2019) 037} [\href{https://arxiv.org/abs/1805.09483}{{\ttfamily
			1805.09483}}].
	
	\bibitem{Bezrukov:2017dyv}
	F.~Bezrukov, M.~Pauly and J.~Rubio, \emph{{On the robustness of the primordial
			power spectrum in renormalized Higgs inflation}},
	\href{https://doi.org/10.1088/1475-7516/2018/02/040}{\emph{JCAP} {\bfseries
			02} (2018) 040} [\href{https://arxiv.org/abs/1706.05007}{{\ttfamily
			1706.05007}}].
	
	\bibitem{Kannike:2017bxn}
	K.~Kannike, L.~Marzola, M.~Raidal and H.~Veerm\"ae, \emph{{Single Field Double
			Inflation and Primordial Black Holes}},
	\href{https://doi.org/10.1088/1475-7516/2017/09/020}{\emph{JCAP} {\bfseries
			09} (2017) 020} [\href{https://arxiv.org/abs/1705.06225}{{\ttfamily
			1705.06225}}].
	
	\bibitem{Lin:2020goi}
	J.~Lin, Q.~Gao, Y.~Gong, Y.~Lu, C.~Zhang and F.~Zhang, \emph{{Primordial black
			holes and secondary gravitational waves from $k$ and $G$ inflation}},
	\href{https://doi.org/10.1103/PhysRevD.101.103515}{\emph{Phys. Rev. D}
		{\bfseries 101} (2020) 103515}
	[\href{https://arxiv.org/abs/2001.05909}{{\ttfamily 2001.05909}}].
	
	\bibitem{Lin:2021vwc}
	J.~Lin, S.~Gao, Y.~Gong, Y.~Lu, Z.~Wang and F.~Zhang, \emph{{Primordial black
			holes and scalar induced gravitational waves from Higgs inflation with
			noncanonical kinetic term}},
	\href{https://doi.org/10.1103/PhysRevD.107.043517}{\emph{Phys. Rev. D}
		{\bfseries 107} (2023) 043517}
	[\href{https://arxiv.org/abs/2111.01362}{{\ttfamily 2111.01362}}].
	
	\bibitem{Gao:2020tsa}
	Q.~Gao, Y.~Gong and Z.~Yi, \emph{{Primordial black holes and secondary
			gravitational waves from natural inflation}},
	\href{https://doi.org/10.1016/j.nuclphysb.2021.115480}{\emph{Nucl. Phys. B}
		{\bfseries 969} (2021) 115480}
	[\href{https://arxiv.org/abs/2012.03856}{{\ttfamily 2012.03856}}].
	
	\bibitem{Gao:2021vxb}
	Q.~Gao, \emph{{Primordial black holes and secondary gravitational waves from
			chaotic inflation}},
	\href{https://doi.org/10.1007/s11433-021-1708-9}{\emph{Sci. China Phys. Mech.
			Astron.} {\bfseries 64} (2021) 280411}
	[\href{https://arxiv.org/abs/2102.07369}{{\ttfamily 2102.07369}}].
	
	\bibitem{Yi:2020kmq}
	Z.~Yi, Y.~Gong, B.~Wang and Z.-h.~Zhu, \emph{{Primordial black holes and
			secondary gravitational waves from the Higgs field}},
	\href{https://doi.org/10.1103/PhysRevD.103.063535}{\emph{Phys. Rev. D}
		{\bfseries 103} (2021) 063535}
	[\href{https://arxiv.org/abs/2007.09957}{{\ttfamily 2007.09957}}].
	
	\bibitem{Yi:2020cut}
	Z.~Yi, Q.~Gao, Y.~Gong and Z.-h.~Zhu, \emph{{Primordial black holes and
			scalar-induced secondary gravitational waves from inflationary models with a
			noncanonical kinetic term}},
	\href{https://doi.org/10.1103/PhysRevD.103.063534}{\emph{Phys. Rev. D}
		{\bfseries 103} (2021) 063534}
	[\href{https://arxiv.org/abs/2011.10606}{{\ttfamily 2011.10606}}].
	
	\bibitem{Yi:2021lxc}
	Z.~Yi and Z.-H.~Zhu, \emph{{NANOGrav signal and LIGO-Virgo primordial black
			holes from the Higgs field}},
	\href{https://doi.org/10.1088/1475-7516/2022/05/046}{\emph{JCAP} {\bfseries
			05} (2022) 046} [\href{https://arxiv.org/abs/2105.01943}{{\ttfamily
			2105.01943}}].
	
	\bibitem{Yi:2022anu}
	Z.~Yi, \emph{{Primordial black holes and scalar-induced gravitational waves
			from the generalized Brans-Dicke theory}},
	\href{https://doi.org/10.1088/1475-7516/2023/03/048}{\emph{JCAP} {\bfseries
			03} (2023) 048} [\href{https://arxiv.org/abs/2206.01039}{{\ttfamily
			2206.01039}}].
	
	\bibitem{Zhang:2020uek}
	F.~Zhang, Y.~Gong, J.~Lin, Y.~Lu and Z.~Yi, \emph{{Primordial non-Gaussianity
			from G-inflation}},
	\href{https://doi.org/10.1088/1475-7516/2021/04/045}{\emph{JCAP} {\bfseries
			04} (2021) 045} [\href{https://arxiv.org/abs/2012.06960}{{\ttfamily
			2012.06960}}].
	
	\bibitem{Pi:2017gih}
	S.~Pi, Y.-l.~Zhang, Q.-G.~Huang and M.~Sasaki, \emph{{Scalaron from
			$R^2$-gravity as a heavy field}},
	\href{https://doi.org/10.1088/1475-7516/2018/05/042}{\emph{JCAP} {\bfseries
			05} (2018) 042} [\href{https://arxiv.org/abs/1712.09896}{{\ttfamily
			1712.09896}}].
	
	\bibitem{Kamenshchik:2018sig}
	A.Y.~Kamenshchik, A.~Tronconi, T.~Vardanyan and G.~Venturi,
	\emph{{Non-Canonical Inflation and Primordial Black Holes Production}},
	\href{https://doi.org/10.1016/j.physletb.2019.02.036}{\emph{Phys. Lett. B}
		{\bfseries 791} (2019) 201}
	[\href{https://arxiv.org/abs/1812.02547}{{\ttfamily 1812.02547}}].
	
	\bibitem{Fu:2019ttf}
	C.~Fu, P.~Wu and H.~Yu, \emph{{Primordial Black Holes from Inflation with
			Nonminimal Derivative Coupling}},
	\href{https://doi.org/10.1103/PhysRevD.100.063532}{\emph{Phys. Rev. D}
		{\bfseries 100} (2019) 063532}
	[\href{https://arxiv.org/abs/1907.05042}{{\ttfamily 1907.05042}}].
	
	\bibitem{Fu:2019vqc}
	C.~Fu, P.~Wu and H.~Yu, \emph{{Scalar induced gravitational waves in inflation
			with gravitationally enhanced friction}},
	\href{https://doi.org/10.1103/PhysRevD.101.023529}{\emph{Phys. Rev. D}
		{\bfseries 101} (2020) 023529}
	[\href{https://arxiv.org/abs/1912.05927}{{\ttfamily 1912.05927}}].
	
	\bibitem{Dalianis:2019vit}
	I.~Dalianis, S.~Karydas and E.~Papantonopoulos, \emph{{Generalized Non-Minimal
			Derivative Coupling: Application to Inflation and Primordial Black Hole
			Production}},
	\href{https://doi.org/10.1088/1475-7516/2020/06/040}{\emph{JCAP} {\bfseries
			06} (2020) 040} [\href{https://arxiv.org/abs/1910.00622}{{\ttfamily
			1910.00622}}].
	
	\bibitem{Gundhi:2020zvb}
	A.~Gundhi and C.F.~Steinwachs, \emph{{Scalaron\textendash{}Higgs inflation
			reloaded: Higgs-dependent scalaron mass and primordial black hole dark
			matter}}, \href{https://doi.org/10.1140/epjc/s10052-021-09225-2}{\emph{Eur.
			Phys. J. C} {\bfseries 81} (2021) 460}
	[\href{https://arxiv.org/abs/2011.09485}{{\ttfamily 2011.09485}}].
	
	\bibitem{Cheong:2019vzl}
	D.Y.~Cheong, S.M.~Lee and S.C.~Park, \emph{{Primordial black holes in
			Higgs-$R^2$ inflation as the whole of dark matter}},
	\href{https://doi.org/10.1088/1475-7516/2021/01/032}{\emph{JCAP} {\bfseries
			01} (2021) 032} [\href{https://arxiv.org/abs/1912.12032}{{\ttfamily
			1912.12032}}].
	
	\bibitem{Zhang:2021rqs}
	F.~Zhang, \emph{{Primordial black holes and scalar induced gravitational waves
			from the E model with a Gauss-Bonnet term}},
	\href{https://doi.org/10.1103/PhysRevD.105.063539}{\emph{Phys. Rev. D}
		{\bfseries 105} (2022) 063539}
	[\href{https://arxiv.org/abs/2112.10516}{{\ttfamily 2112.10516}}].
	
	\bibitem{Zhang:2021vak}
	F.~Zhang, J.~Lin and Y.~Lu, \emph{{Double-peaked inflation model: Scalar
			induced gravitational waves and primordial-black-hole suppression from
			primordial non-Gaussianity}},
	\href{https://doi.org/10.1103/PhysRevD.104.063515}{\emph{Phys. Rev. D}
		{\bfseries 104} (2021) 063515}
	[\href{https://arxiv.org/abs/2106.10792}{{\ttfamily 2106.10792}}].
	
	\bibitem{Kawai:2021edk}
	S.~Kawai and J.~Kim, \emph{{Primordial black holes from Gauss-Bonnet-corrected
			single field inflation}},
	\href{https://doi.org/10.1103/PhysRevD.104.083545}{\emph{Phys. Rev. D}
		{\bfseries 104} (2021) 083545}
	[\href{https://arxiv.org/abs/2108.01340}{{\ttfamily 2108.01340}}].
	
	\bibitem{Cai:2021wzd}
	R.-G.~Cai, C.~Chen and C.~Fu, \emph{{Primordial black holes and stochastic
			gravitational wave background from inflation with a noncanonical spectator
			field}}, \href{https://doi.org/10.1103/PhysRevD.104.083537}{\emph{Phys. Rev.
			D} {\bfseries 104} (2021) 083537}
	[\href{https://arxiv.org/abs/2108.03422}{{\ttfamily 2108.03422}}].
	
	\bibitem{Chen:2021nio}
	P.~Chen, S.~Koh and G.~Tumurtushaa, \emph{{Primordial black holes and induced
			gravitational waves from inflation in the Horndeski theory of gravity}},
	\href{https://arxiv.org/abs/2107.08638}{{\ttfamily 2107.08638}}.
	
	\bibitem{Zheng:2021vda}
	R.~Zheng, J.~Shi and T.~Qiu, \emph{{On primordial black holes and secondary
			gravitational waves generated from inflation with solo/multi-bumpy potential
			*}}, \href{https://doi.org/10.1088/1674-1137/ac42bd}{\emph{Chin. Phys. C}
		{\bfseries 46} (2022) 045103}
	[\href{https://arxiv.org/abs/2106.04303}{{\ttfamily 2106.04303}}].
	
	\bibitem{Karam:2022nym}
	A.~Karam, N.~Koivunen, E.~Tomberg, V.~Vaskonen and H.~Veerm\"ae, \emph{{Anatomy
			of single-field inflationary models for primordial black holes}},
	\href{https://doi.org/10.1088/1475-7516/2023/03/013}{\emph{JCAP} {\bfseries
			03} (2023) 013} [\href{https://arxiv.org/abs/2205.13540}{{\ttfamily
			2205.13540}}].
	
	\bibitem{Ashoorioon:2019xqc}
	A.~Ashoorioon, A.~Rostami and J.T.~Firouzjaee, \emph{{EFT compatible PBHs:
			effective spawning of the seeds for primordial black holes during
			inflation}}, \href{https://doi.org/10.1007/JHEP07(2021)087}{\emph{JHEP}
		{\bfseries 07} (2021) 087}
	[\href{https://arxiv.org/abs/1912.13326}{{\ttfamily 1912.13326}}].
	
	\bibitem{Danzmann:1997hm}
	K.~Danzmann, \emph{{LISA: An ESA cornerstone mission for a gravitational wave
			observatory}}, \href{https://doi.org/10.1088/0264-9381/14/6/002}{\emph{Class.
			Quant. Grav.} {\bfseries 14} (1997) 1399}.
	
	\bibitem{Audley:2017drz}
	{\scshape LISA} collaboration, \emph{{Laser Interferometer Space Antenna}},
	\href{https://arxiv.org/abs/1702.00786}{{\ttfamily 1702.00786}}.
	
	\bibitem{Hu:2017mde}
	W.-R.~Hu and Y.-L.~Wu, \emph{{The Taiji Program in Space for gravitational wave
			physics and the nature of gravity}},
	\href{https://doi.org/10.1093/nsr/nwx116}{\emph{Natl. Sci. Rev.} {\bfseries
			4} (2017) 685}.
	
	\bibitem{Luo:2015ght}
	{\scshape TianQin} collaboration, \emph{{TianQin: a space-borne gravitational
			wave detector}},
	\href{https://doi.org/10.1088/0264-9381/33/3/035010}{\emph{Class. Quant.
			Grav.} {\bfseries 33} (2016) 035010}
	[\href{https://arxiv.org/abs/1512.02076}{{\ttfamily 1512.02076}}].
	
	\bibitem{Gong:2021gvw}
	Y.~Gong, J.~Luo and B.~Wang, \emph{{Concepts and status of Chinese space
			gravitational wave detection projects}},
	\href{https://doi.org/10.1038/s41550-021-01480-3}{\emph{Nature Astron.}
		{\bfseries 5} (2021) 881} [\href{https://arxiv.org/abs/2109.07442}{{\ttfamily
			2109.07442}}].
	
	\bibitem{Kawamura:2011zz}
	S.~Kawamura et~al., \emph{{The Japanese space gravitational wave antenna:
			DECIGO}}, \href{https://doi.org/10.1088/0264-9381/28/9/094011}{\emph{Class.
			Quant. Grav.} {\bfseries 28} (2011) 094011}.
	
	\bibitem{Inomata:2018epa}
	K.~Inomata and T.~Nakama, \emph{{Gravitational waves induced by scalar
			perturbations as probes of the small-scale primordial spectrum}},
	\href{https://doi.org/10.1103/PhysRevD.99.043511}{\emph{Phys. Rev. D}
		{\bfseries 99} (2019) 043511}
	[\href{https://arxiv.org/abs/1812.00674}{{\ttfamily 1812.00674}}].
	
	\bibitem{Yi:2022ymw}
	Z.~Yi and Q.~Fei, \emph{{Constraints on primordial curvature spectrum from
			primordial black holes and scalar-induced gravitational waves}},
	\href{https://doi.org/10.1140/epjc/s10052-023-11233-3}{\emph{Eur. Phys. J. C}
		{\bfseries 83} (2023) 82} [\href{https://arxiv.org/abs/2210.03641}{{\ttfamily
			2210.03641}}].
	
	\bibitem{Lu:2019sti}
	Y.~Lu, Y.~Gong, Z.~Yi and F.~Zhang, \emph{{Constraints on primordial curvature
			perturbations from primordial black hole dark matter and secondary
			gravitational waves}},
	\href{https://doi.org/10.1088/1475-7516/2019/12/031}{\emph{JCAP} {\bfseries
			12} (2019) 031} [\href{https://arxiv.org/abs/1907.11896}{{\ttfamily
			1907.11896}}].
	
	\bibitem{Inomata:2016rbd}
	K.~Inomata, M.~Kawasaki, K.~Mukaida, Y.~Tada and T.T.~Yanagida,
	\emph{{Inflationary primordial black holes for the LIGO gravitational wave
			events and pulsar timing array experiments}},
	\href{https://doi.org/10.1103/PhysRevD.95.123510}{\emph{Phys. Rev. D}
		{\bfseries 95} (2017) 123510}
	[\href{https://arxiv.org/abs/1611.06130}{{\ttfamily 1611.06130}}].
	
	\bibitem{Espinosa:2018eve}
	J.R.~Espinosa, D.~Racco and A.~Riotto, \emph{{A Cosmological Signature of the
			SM Higgs Instability: Gravitational Waves}},
	\href{https://doi.org/10.1088/1475-7516/2018/09/012}{\emph{JCAP} {\bfseries
			09} (2018) 012} [\href{https://arxiv.org/abs/1804.07732}{{\ttfamily
			1804.07732}}].
	
	\bibitem{Ashton:2018jfp}
	G.~Ashton et~al., \emph{{BILBY: A user-friendly Bayesian inference library for
			gravitational-wave astronomy}},
	\href{https://doi.org/10.3847/1538-4365/ab06fc}{\emph{Astrophys. J. Suppl.}
		{\bfseries 241} (2019) 27}
	[\href{https://arxiv.org/abs/1811.02042}{{\ttfamily 1811.02042}}].
	
	\bibitem{NestedSampling}
	J.~Skilling, \emph{{Nested Sampling}},
	\href{https://doi.org/10.1063/1.1835238}{\emph{AIP Conf. Proc.} {\bfseries
			735} (2004) 395}.
	
	\bibitem{Moore:2021ibq}
	C.J.~Moore and A.~Vecchio, \emph{{Ultra-low-frequency gravitational waves from
			cosmological and astrophysical processes}},
	\href{https://doi.org/10.1038/s41550-021-01489-8}{\emph{Nature Astron.}
		{\bfseries 5} (2021) 1268}
	[\href{https://arxiv.org/abs/2104.15130}{{\ttfamily 2104.15130}}].
	
	\bibitem{Inomata:2016uip}
	K.~Inomata, M.~Kawasaki and Y.~Tada, \emph{{Revisiting constraints on small
			scale perturbations from big-bang nucleosynthesis}},
	\href{https://doi.org/10.1103/PhysRevD.94.043527}{\emph{Phys. Rev. D}
		{\bfseries 94} (2016) 043527}
	[\href{https://arxiv.org/abs/1605.04646}{{\ttfamily 1605.04646}}].
	
	\bibitem{Fixsen:1996nj}
	D.J.~Fixsen, E.S.~Cheng, J.M.~Gales, J.C.~Mather, R.A.~Shafer and E.L.~Wright,
	\emph{{The Cosmic Microwave Background spectrum from the full COBE FIRAS data
			set}}, \href{https://doi.org/10.1086/178173}{\emph{Astrophys. J.} {\bfseries
			473} (1996) 576} [\href{https://arxiv.org/abs/astro-ph/9605054}{{\ttfamily
			astro-ph/9605054}}].
	
\end{thebibliography}

\providecommand{\href}[2]{#2}\begingroup\raggedright\endgroup

\end{document}